\documentclass[manuscript, screen, nonacm]{acmart}

\usepackage[yyyymmdd]{datetime}

\usepackage{acmart-taps}
\usepackage{eurosym}
\usepackage{tabularx}
\usepackage{dcolumn} 
\newcolumntype{d}[1]{D{.}{.}{#1}}
\usepackage{xcolor}
\usepackage{amsmath}
\usepackage{enumitem}
\usepackage{caption}
\usepackage{subcaption}
\usepackage{CJKutf8}

\newcommand{\cn}[1]{\begin{CJK*}{UTF8}{gbsn}#1\end{CJK*}}

\AtBeginDocument{%
  }

\authorsaddresses{}

\begin{document}

\title{A Systematic and Validated Translation of the Perceived Empathy of Technology Scale to Chinese}

\settopmatter{authorsperrow=3}

\author{Matthias Schmidmaier}
\orcid{0009-0006-8609-8716}
\affiliation{%
  \institution{LMU Munich}
  \city{Munich}
  \country{Germany}}
\affiliation{%
  \institution{TU Dortmund University}
  \city{Dortmund}
  \country{Germany}}
\email{matt@schmidmaier.org}

\author{Jingyao Zheng}
\orcid{0000-0001-8920-308X}
\affiliation{%
  \institution{Hong Kong Polytechnic University}
  \city{Hong Kong}
  \country{China}}
\email{jingyao.zheng@connect.polyu.hk}

\author{Jie Song}
\orcid{0009-0004-8515-4195}
\affiliation{%
  \institution{LMU Munich}
  \city{Munich}
  \country{Germany}}
\email{jie.song@campus.lmu.de}

\author{Lik-Hang Lee}
\orcid{0000-0003-1361-1612}
\affiliation{%
  \institution{Hong Kong Polytechnic University}
  \city{Hong Kong}
  \country{China}}
\email{lik-hang.lee@polyu.edu.hk}

\author{Sven Mayer}
\orcid{0000-0001-5462-8782}
\affiliation{%
  \institution{TU Dortmund University}
  \city{Dortmund}
  \country{Germany}}
\email{info@sven-mayer.com}

\renewcommand{\shortauthors}{Schmidmaier et al.}

\begin{abstract}
The Perceived Empathy of Technology Scale (PETS) allows evaluation of empathic behavior in intelligent systems, a concept increasingly applied in human-AI interaction, particularly in emotional and mental health support applications.
To support intercultural research and development, PETS is already available in multiple languages, as research suggests cultural and language-based differences in the expression and perception of empathy.
Given the substantial Chinese-speaking population and the growing focus on cross-cultural HCI research, we present a validated translation of PETS into Putonghua (\cn{普通话}, Mandarin) with Simplified Chinese (\cn{简体中文}).
Based on established back-translation methodologies as used in previous PETS translations, our process included multiple independent translations, expert panel reviews, and empirical testing with a total of $N=400$ participants from an English and a Chinese sample.
Through confirmatory factor analysis and invariance testing we show that the Chinese PETS retains the original two-factor structure with high reliability, with support for configural, metric, and scalar invariance.
Beyond validation, empathic system behavior was rated equivalently across languages, while non-empathic behavior drew substantially higher ratings from Chinese participants, suggesting its absence is read less consistently across cultures than its presence.
This validated Chinese version enables researchers and practitioners to assess empathy perceptions among Chinese-speaking users, supporting culturally appropriate empathic technology development.
\end{abstract}



\maketitle

\noindent{\textbf{Preprint of the manuscript submitted to ICHEC 2026}.
\\Licensed under \href{https://creativecommons.org/licenses/by/4.0/}{CC BY 4.0}. 
This preprint has not undergone peer review (when applicable) or any post-submission improvements or corrections.
}
\medskip

\begin{table*}[t]

\begin{minipage}{0.48\linewidth}
\caption{The validated Chinese PETS. Use with randomized 101-point sliders from \cn{非常不同意} to \cn{非常同意}. 
}
\label{tab:pets}
\centering
\small
\begin{tabularx}{\linewidth}{rX}
\toprule
\textbf{PETS-ER} & \cn{\textbf{情感响应能力}}\\
\midrule
 E1 & \cn{该系统考虑到了我的心理状态。}  \\
 E2 & \cn{该系统表现出情感智能。} \\
 E3 & \cn{该系统表达出了情感。 }\\
 E4 & \cn{该系统能与我共情。}\\
 E5 & \cn{该系统对我表现出了兴趣。}\\
 E6 & \cn{该系统在我应对情感状况时给予了支持。}\\
 \\
\midrule
\textbf{PETS-UT} & \cn{\textbf{理解与信任}} \\
\midrule
 U1 & \cn{该系统理解了我的目标。}\\
 U2 & \cn{该系统理解了我的需求。}\\
 U3 & \cn{我对该系统产生了信任。}\\
 U4 & \cn{该系统理解了我的意图。}\\
\bottomrule
\end{tabularx}
\end{minipage}
\hfill
\begin{minipage}{.48\linewidth}
\caption{The original English PETS \cite{schmidmaier2024perceived}. Use with randomized 101-point sliders from ``strongly disagree'' to ``strongly agree''.}
\label{tab:pets_en}
\centering
\small
\begin{tabularx}{\linewidth}{rX}
\toprule
\textbf{PETS-ER} & \textbf{Emotional Responsiveness}\\
\midrule
 E1 & The system considered my mental state.  \\
 E2 & The system seemed emotionally intelligent. \\
 E3 & The system expressed emotions. \\
 E4 & The system sympathized with me.\\
 E5 & The system showed interest in me. \\
 E6 & The system supported me in coping with an emotional
situation. \\
\midrule
\textbf{PETS-UT} & \textbf{Understanding and Trust} \\
\midrule
 U1 & The system understood my goals. \\
 U2 & The system understood my needs. \\
 U3 & I trusted the system. \\
 U4 & The system understood my intentions. \\
\bottomrule
\end{tabularx}
\end{minipage}

\vspace{4pt}
{\footnotesize\raggedright
\textit{Note.} Freely available at \url{https://perceived-empathy-of-technology-scale.com} under CC BY 4.0.\par}

\end{table*}


\section{Introduction}
Intelligent artificial agents are increasingly required to express empathic behavior, especially in the context of well-being, digital health support, and social care~\cite{Park2022-ng,Jeong2023-hq,Paiva2017-qb}, where empathy can affect user trust, engagement~\cite{yoon2021ai, plank1996short, Johanson2023-pc}, and emotional support~\cite{Bae_Brandtzaeg2021-jc,Bickmore2005-rg,De_Gennaro2019-rl}.
Technological advances in generative AI and affective computing now enable systems to process and express emotional states~\cite{Park2022-ng,Elyoseph2023-ip,Liu2018-yh,Daher2020-rv} in a way that, in some cases, even exceeds human empathy abilities~\cite{Ayers2023-pq,Liu2024-bm,Yonatan-Leus2025-fd}.
Evaluating how users actually perceive that behavior therefore becomes a prerequisite for designing such systems responsibly.
To evaluate the empathic capabilities of artificial systems, \citet{schmidmaier2024perceived} introduced the Perceived Empathy of Technology Scale (PETS).
The PETS was developed and validated in English, and \citet{schmidmaier2025systematic} subsequently published a German version, motivated by evidence that emotional perception and empathic response vary with cultural background~\cite{Pavlenko2012-az,Ward2025-or,Keysar2012-la} and native language~\cite{Dewaele2010-ra,Schmitt2020-pe}.
Such variation is particularly pronounced in Chinese contexts, where empathy relies heavily on nonverbal cues and contextual sensitivity rather than explicit emotional expression, reflecting Confucian values of emotional moderation and face-conscious interaction within relationship networks (``guanxi'', \cn{关系})~\cite{Guo2024EmpathyAI,liu2024understanding}.
Chinese individuals also tend to experience greater self-focused distress when confronted with others' suffering and to prefer practical advice over direct emotional validation, contrasting with Western patterns of other-oriented empathic concern~\cite{cassels2010role,trommsdorff2007sympathy}.
Yet no instrument for perceived empathy of technology has been validated for Chinese-speaking populations.
Given how widely empathic systems are deployed in China, this leaves a substantial body of work without a comparable measure.
We therefore developed and validated a Chinese version of the PETS, targeting Putonghua (\cn{普通话}) and Simplified Chinese (\cn{简体中文}).
We closely followed the translation and validation process applied by \citet{schmidmaier2025systematic}, which builds on work by \citet{Brislin1970-by,Klotz2023-ww} and \citet{Jones2001-wa}.
Confirmatory factor analysis established measurement equivalence with the English PETS across configural, metric, and scalar levels, with high reliability throughout.
Unlike the German translation, where the trust item U3 loaded lower than in English~\cite{schmidmaier2025systematic}, the Chinese version showed a stronger loading for that item.
Comparing scale behavior across languages further revealed an asymmetry, as ratings of empathic system behavior were equivalent across samples while non-empathic behavior was rated substantially higher by Chinese participants, which may indicate that the absence of empathy is interpreted less consistently across cultures than its presence.
This work thus contributes a validated Chinese PETS for cross-cultural research on empathic technologies, together with evidence that cultural differences in perceived empathy surface where empathic behavior is absent rather than where it is present.
Additional information on how to apply the PETS can be found at \url{ https://perceived-empathy-of-technology-scale.com}.

\section{Related Work}
In this section, we introduce empathy and its role in HCI, including cross-cultural aspects and Chinese applications, as well as the PETS and the foundations of scientific back-translation.

\subsection{Empathic Systems}
Empathy is defined mainly as a multidimensional construct encompassing both cognitive understanding of emotional states and affective emotional responses~\cite{davis1983measuring,Batson2009-uu,Cuff2016-fg}.
It plays a fundamental role in human interaction and is increasingly explored in the context of artificial systems such as chatbots~\cite{Liu2025-oc,Morris2018-sj,Cuadra2024-yt}, virtual agents~\cite{Parmar2022-af}, or social robots~\cite{Paiva2017-qb,Park2022-ng}.
For example, embedding empathic behavior in such systems has been found to increase user engagement~\cite{Seitz2024-is,Charrier2018-iu} and trust~\cite{Seitz2024-is,Cramer2010-kj,Johanson2023-pc}, provide emotional support~\cite{Bae_Brandtzaeg2021-jc,Erel2022-dm}, counter social exclusion~\cite{De_Gennaro2019-rl}, or assist in mental health support~\cite{Jeong2023-hq,Xie2024-hk,Collins2024-vp}.
Furthermore, modern LLM-based conversational systems have demonstrated capabilities comparable to those of humans in terms of general emotional awareness~\cite{Elyoseph2023-ip} and the generation of empathic responses, sometimes even surpassing them, for example, in the contexts of medical communication~\cite{Ayers2023-pq,Liu2024-bm,Webb2023-yg} or psychotherapy~\cite{Yonatan-Leus2025-fd}.
Particularly in healthcare, however, the use of empathic agents raises ethical challenges with regard to data protection~\cite{Bae_Brandtzaeg2021-jc,De-Choudhury2023-ws,Cuadra2024-yt}, safety, and quality of care~\cite{Dekeyser2009-qz,De-Choudhury2023-ws,Prescott2023-bw}, as systems may give incorrect advice or exhibit inherent biases~\cite{De-Choudhury2023-ws,Cabrera2023-bb}.
Deploying such systems responsibly therefore depends on being able to evaluate how users perceive them, which in turn requires measurement instruments that function in the language of use.
Prominent Chinese applications in this context include Microsoft's empathic agent XiaoIce (\cn{小冰}) and Baidu's Lingyi Zhihui (\cn{灵医智惠}), a healthcare-oriented conversational agent based on Wenxin Yiyan (\cn{文心一言}), a general-purpose conversational model that is increasingly integrated into education, customer service, and government portals.
Further examples are Doubao (\cn{豆包}, ByteDance), a conversational AI for empathic, emotionally expressive dialogue in consumer and enterprise settings, and Xiaomi's XiaoAI smart voice assistant, which was found to provide companionship and alleviate loneliness, especially for elderly users~\cite{chung2020using}.
Systems of this kind reach large user bases and could thus profit from an instrument that captures perceived empathy in the language of their users.

\subsection{Empathy in Chinese Culture}
Also in Chinese culture, empathy, translated as \cn{共情} or \cn{移情}, is conceptualized as a dynamic process that combines cognitive understanding (\cn{认知共情}) with emotional resonance (\cn{情感共情})~\cite{Guo2024EmpathyAI,ho2004guilt}.
However, it is characterized by distinct cultural and relational patterns that differ from Western conceptualizations.
In the Chinese context, empathy is shaped by collectivist values and Confucian traditions, particularly the Confucian ``Doctrine of the Mean'' (\cn{中庸}) philosophy, which encourages moderation in emotional display and promotes indirect communication to preserve harmony and face~\cite{bond1993emotions,frijda2007emotion,ho2004guilt,zhu2007neural}.
As a result, empathic interactions often rely more on subtle nonverbal cues such as tone, gesture, and situational context than on open verbalization of personal feelings~\cite{ho2004guilt,siu2005validation}.
This tendency resonates with Fei's framework of ``chaxu geju'' (\cn{差序格局}), in which individuals define themselves relationally, incorporating family members, friends, and colleagues into the self-concept rather than maintaining strict individual boundaries~\cite{yan2020individualization,marong2007hierarch,markus1991culture,zhu2002experimental}.
In line with this relational orientation, Chinese empathy is also linked to implicit theories of emotion that emphasize the possibility of regulating and transforming one's emotional states, reflecting cultural ideals of self-discipline and social harmony~\cite{kornilova2021cross}.
While this relational orientation and emphasis on emotional regulation can strengthen in-group empathy, it may reduce perspective-taking toward others and hinder empathic accuracy~\cite{chopik2017differences,atkins2016culture,cheon2010empathy}.
Cross-cultural research supports these patterns, showing that East Asian participants, including Chinese, often report higher levels of self-focused personal distress but lower levels of other-oriented empathic concern compared to Western participants~\cite{cassels2010role,trommsdorff2007sympathy}.
These findings suggest that empathy in Chinese contexts is more relationally embedded and less individualistic in orientation.
Such culturally grounded dynamics are also reflected in the design of empathic systems in China, where conversational agents and digital assistants often employ indirect communication, context sensitivity, and relationally attuned strategies to align with users' expectations~\cite{zhou2020design,liu2023can,yu2022data}.
How these patterns translate to judgments of artificial systems is not yet established, as the literature describes empathy between people rather than empathy attributed to technology, leaving open whether the same relational orientation shapes how users read empathic and non-empathic system behavior.

\subsection{The Cross-Cultural PETS}
The Perceived Empathy of Technology Scale (PETS), introduced in 2024, enables a modern, validated measurement of how users perceive the empathic behavior of technological systems~\cite{schmidmaier2024perceived}.
As shown in \autoref{tab:pets_en}, the original PETS comprises two factors -- Emotional Responsiveness (PETS-ER) and Understanding and Trust (PETS-UT) -- with a total of ten items, where PETS-ER assesses emotional understanding and support while PETS-UT evaluates system understanding of user goals and trustworthiness~\cite{schmidmaier2024perceived}.
According to the authors, PETS should be applied after users interacted with a system, using 101-point sliders ranging from ``strongly disagree'' to ``strongly agree'' and randomized item order~\cite{schmidmaier2024perceived}.
The scale was originally developed and validated in English.
In a follow-up publication, \citet{schmidmaier2025systematic} introduced a German version of the PETS to enable research in users' native language and to facilitate cross-cultural evaluation.
For translation and validation, the authors followed a refined back-translation process~\cite{Brislin1970-by,Jones2001-wa,Klotz2023-ww} and demonstrated configural, metric, and scalar measurement invariance between the translated and the original scale~\cite{schmidmaier2025systematic}.
They also observed a reduced loading for the trust item U3, which they attribute to cultural and individual variation in how trust in a system is understood.
Furthermore, their CFA results showed slightly elevated RMSEA values for the German PETS, which the authors attributed to the simplicity of the two-factor model with relatively few degrees of freedom~\cite{Kenny2003-hn}, pointing to excellent CFI, TLI, and SRMR scores as evidence of an overall good model fit.
In this work, we follow up on this approach by developing and validating a Chinese version of the PETS, aiming to provide a culturally appropriate measurement tool for Chinese contexts while investigating potential cross-cultural differences in empathy perception and scale structure.

\subsection{Translation for Cross-Cultural Research}
\label{ss:rw-translation}
As described in \cite{schmidmaier2024perceived}, the back-translation procedure, originally introduced by \citet{Brislin1970-by}, established a foundational approach for cross-cultural research.
It involved four iterative steps: translation, blind back-translation, comparison of the two versions, and revision.
In addition to the method itself, \citet{Brislin1970-by} proposed five criteria for evaluating translation quality and demonstrated that factors such as the type of content, the level of linguistic difficulty, and the degree of similarity between languages significantly shape translation outcomes.
Building on this work, \citet{Jones2001-wa} introduced refinements that addressed some of the limitations of Brislin's sequential approach.
Their model emphasized simultaneous independent forward translations by multiple bilingual experts, which were then reconciled through collaborative group discussions.
The revised process also included blind back-translation and cross-lingual validation with bilingual participants, allowing researchers to detect subtle semantic differences that might otherwise be overlooked.
Despite its widespread adoption, back-translation practices have been criticized for insufficient methodological rigor.
Reviewing more than 300 studies, \citet{Klotz2023-ww} found that although back-translation remains the dominant strategy in scale adaptation, only a small fraction of studies reported systematic pretesting, and even fewer presented quantitative evidence for equivalence.
To strengthen validity, they recommend supplementing back-translation with qualitative techniques (e.g., expert committee reviews, random-probe interviews) and statistical approaches such as confirmatory factor analysis to establish measurement invariance.
\citet{schmidmaier2025systematic} applied these recommendations for their German PETS translation process, integrating multiple forward and back-translations, expert panel discussions, and final statistical validation through confirmatory factor analysis.
The authors also highlight the potential role of AI-based translation in streamlining the initial translation, though they stress that human expertise remains indispensable for ensuring conceptual accuracy and for addressing complex constructs, particularly in domains such as empathy research.

\section{Methodology}
\label{s:methodology}
\begin{figure*}[t]
    \centering
    \includegraphics[width=\linewidth]{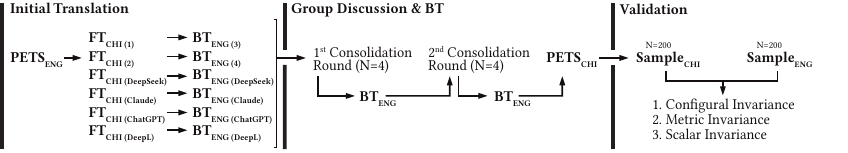}
    \caption{PETS translation process from English (ENG) to Chinese (CHI) with multiple forward (FT) and back translation (BT) steps from human translators, and different AI models (DeepSeek, Claude, ChatGPT, DeepL). Two collaborative consolidation rounds resulted in a consolidated Chinese PETS, which was then validated with Confirmatory Factor Analysis (CFA).}
    \label{fig:process} 
\end{figure*}

We developed the Chinese PETS with the same systematic procedure that was previously applied to the German version~\cite{schmidmaier2025systematic}, which pairs refined back-translation~\cite{Jones2001-wa} with statistical validation~\cite{Klotz2023-ww}.
The target variety was Simplified Chinese (\cn{简体中文}) as used in mainland China.
\autoref{fig:process} outlines the three phases.
First, several human translators and AI systems worked independently, each contributing forward and back-translations (\autoref{ss:translation}).
With these drafts we then conducted a moderated panel of bilingual experts (\autoref{ss:group_discussion}), who merged them into one Chinese version that we had back-translated so that it could be checked against the original items.
Finally, we ran confirmatory factor analysis and tests of measurement invariance on parallel Chinese and English samples (\autoref{s:validation}).


\section{Translation} 
As described in \autoref{s:methodology}, the first two phases of our process resulted in scale translation, involving initial forward and back-translations, expert group discussion for consolidation, and final back-translations.

\subsection{Initial Translation}
\label{ss:translation}
We commissioned two independent forward and two back-translations from bilingual translators recruited through the online platform \href{https://fiverr.com}{Fiverr}.
Compensation per translation ranged from $7$ \euro to $10$ \euro, depending on individual experience and asking price.
The translators worked independently and received only the original PETS items, a brief contextual introduction, and instructions to translate into Simplified Chinese (\cn{简体中文}), using a neutral tone and avoiding regional idioms.
In parallel, we generated four more translations with DeepL (21-3-25), Claude Sonnet (3.7), ChatGPT (gpt-4o), and DeepSeek (v3).
This initial translation phase resulted in six forward and six back-translations.

\subsection{Group Discussion}
\label{ss:group_discussion}
In the second phase, we consolidated the initial translations into a single Chinese version.
It proceeded as an iterative loop of expert discussion and back-translation, repeated until the reconstructed English no longer diverged from the source items.

\providecommand{\sqf}{\mathord{\scalebox{0.9}{$\blacksquare$}}}
\providecommand{\sqe}{\mathord{\scalebox{0.9}{$\square$}}}

\begin{table*}[t]
\caption{Demographic background and self-assessed language skills and expertise in related fields of the participants of the initial discussion (P1--P4) and the second consolidation round (P5--P8).}
\label{tab:experts_rating_inverted}
\centering
\small
\setlength{\tabcolsep}{4pt}
\renewcommand{\arraystretch}{1}
\begin{tabularx}{\linewidth}{
Xcccc@{\hskip 18pt}cccc@{\hskip 12pt}d{2.2} d{1.2}
}
\toprule
\ 
& \textbf{P1} & \textbf{P2} & \textbf{P3} & \textbf{P4} 
& \textbf{P5} & \textbf{P6} & \textbf{P7} & \textbf{P8} & \multicolumn{1}{c}{\textbf{Avg}} & \multicolumn{1}{c}{\textbf{SD}} \\
\midrule
Age & 25 & 32 & 29 & 30 & 24 & 26 & 23 & 25 & 26.75 & 2.99\\
Gender & male & male & male & female & female & male & female & male & - & -\\
Degree & 
MA & MA & PhD & MA & MA & MA & BA & MA & - & -\\
\midrule
\multicolumn{11}{l}{\footnotesize\textit{Expertise assessment}} \\
Psychology 
& $\sqf$$\sqf$$\sqf$$\sqe$$\sqe$ 
& $\sqf$$\sqe$$\sqe$$\sqe$$\sqe$ 
& $\sqf$$\sqf$$\sqf$$\sqe$$\sqe$ 
& $\sqf$$\sqe$$\sqe$$\sqe$$\sqe$ 
& $\sqf$$\sqf$$\sqf$$\sqe$$\sqe$ 
& $\sqf$$\sqf$$\sqf$$\sqe$$\sqe$ 
& $\sqf$$\sqf$$\sqf$$\sqf$$\sqf$ 
& $\sqf$$\sqf$$\sqe$$\sqe$$\sqe$ 
& 2.62 & 1.22\\
HCI 
& $\sqf$$\sqf$$\sqf$$\sqf$$\sqe$ & $\sqf$$\sqe$$\sqe$$\sqe$$\sqe$ & $\sqf$$\sqf$$\sqe$$\sqe$$\sqe$ & $\sqf$$\sqf$$\sqf$$\sqf$$\sqe$ &  $\sqf$$\sqf$$\sqf$$\sqf$$\sqe$ & $\sqf$$\sqf$$\sqf$$\sqf$$\sqf$ & $\sqf$$\sqf$$\sqf$$\sqf$$\sqe$ & $\sqf$$\sqf$$\sqf$$\sqf$$\sqe$ & 3.50 & 1.22\\

Affective systems 
& $\sqf$$\sqf$$\sqf$$\sqf$$\sqe$ & $\sqf$$\sqf$$\sqe$$\sqe$$\sqe$ & $\sqf$$\sqf$$\sqf$$\sqf$$\sqe$ & $\sqf$$\sqf$$\sqe$$\sqe$$\sqe$ & $\sqf$$\sqf$$\sqf$$\sqf$$\sqe$ & $\sqf$$\sqf$$\sqf$$\sqf$$\sqf$ & $\sqf$$\sqf$$\sqf$$\sqf$$\sqf$ & $\sqf$$\sqf$$\sqf$$\sqf$$\sqe$ & 3.75 & 1.09\\

ESB theories & $\sqf$$\sqf$$\sqf$$\sqf$$\sqe$ & $\sqf$$\sqf$$\sqe$$\sqe$$\sqe$ & $\sqf$$\sqf$$\sqf$$\sqf$$\sqe$ & $\sqf$$\sqf$$\sqf$$\sqe$$\sqe$ & $\sqf$$\sqf$$\sqf$$\sqe$$\sqe$ & $\sqf$$\sqf$$\sqf$$\sqf$$\sqf$ & $\sqf$$\sqf$$\sqf$$\sqf$$\sqf$ & $\sqf$$\sqf$$\sqf$$\sqe$$\sqe$ & 3.62 & .99\\

Empathy theories & $\sqf$$\sqf$$\sqe$$\sqe$$\sqe$ & $\sqf$$\sqe$$\sqe$$\sqe$$\sqe$ & $\sqf$$\sqf$$\sqf$$\sqf$$\sqe$ & $\sqf$$\sqf$$\sqe$$\sqe$$\sqe$ & $\sqf$$\sqf$$\sqf$$\sqf$$\sqf$ & $\sqf$$\sqf$$\sqf$$\sqf$$\sqf$ & $\sqf$$\sqf$$\sqf$$\sqf$$\sqf$ & $\sqf$$\sqf$$\sqf$$\sqf$$\sqe$ & 3.50 & 1.50\\

Scale application & $\sqf$$\sqf$$\sqf$$\sqf$$\sqe$ & $\sqf$$\sqf$$\sqf$$\sqf$$\sqe$ & $\sqf$$\sqf$$\sqf$$\sqf$$\sqe$ & $\sqf$$\sqf$$\sqf$$\sqf$$\sqe$ & $\sqf$$\sqf$$\sqf$$\sqf$$\sqe$ & $\sqf$$\sqf$$\sqf$$\sqf$$\sqf$ & $\sqf$$\sqf$$\sqf$$\sqf$$\sqf$ & $\sqf$$\sqf$$\sqe$$\sqe$$\sqe$ & 4.00 & .87\\

Scale development & $\sqf$$\sqf$$\sqe$$\sqe$$\sqe$ & $\sqf$$\sqf$$\sqf$$\sqf$$\sqe$ & $\sqf$$\sqe$$\sqe$$\sqe$$\sqe$ & $\sqf$$\sqf$$\sqe$$\sqe$$\sqe$ & $\sqf$$\sqf$$\sqf$$\sqf$$\sqe$ & $\sqf$$\sqf$$\sqf$$\sqf$$\sqe$ & $\sqf$$\sqf$$\sqf$$\sqf$$\sqf$ & $\sqf$$\sqf$$\sqe$$\sqe$$\sqe$ & 3.00 & 1.32\\

\midrule
\multicolumn{11}{l}{\footnotesize\textit{Language skill assessment}} \\
English Reading & $\sqf$$\sqf$$\sqf$$\sqf$$\sqe$ & $\sqf$$\sqf$$\sqf$$\sqf$$\sqe$ & $\sqf$$\sqf$$\sqf$$\sqf$$\sqe$ & $\sqf$$\sqf$$\sqf$$\sqf$$\sqe$ & $\sqf$$\sqf$$\sqf$$\sqf$$\sqf$ & $\sqf$$\sqf$$\sqf$$\sqf$$\sqe$ & $\sqf$$\sqf$$\sqf$$\sqf$$\sqe$ & $\sqf$$\sqf$$\sqf$$\sqf$$\sqf$ & 4.25 & .43\\
English Writing & $\sqf$$\sqf$$\sqf$$\sqf$$\sqe$ & $\sqf$$\sqf$$\sqf$$\sqf$$\sqe$ & $\sqf$$\sqf$$\sqf$$\sqf$$\sqe$ & $\sqf$$\sqf$$\sqf$$\sqf$$\sqe$ & $\sqf$$\sqf$$\sqf$$\sqf$$\sqf$ & $\sqf$$\sqf$$\sqf$$\sqf$$\sqe$ & $\sqf$$\sqf$$\sqf$$\sqe$$\sqe$ & $\sqf$$\sqf$$\sqf$$\sqf$$\sqf$ & 4.12 & .60\\

Chinese Reading & $\sqf$$\sqf$$\sqf$$\sqf$$\sqf$ & $\sqf$$\sqf$$\sqf$$\sqf$$\sqf$ & $\sqf$$\sqf$$\sqf$$\sqf$$\sqe$ & $\sqf$$\sqf$$\sqf$$\sqf$$\sqf$ & $\sqf$$\sqf$$\sqf$$\sqf$$\sqf$ & $\sqf$$\sqf$$\sqf$$\sqf$$\sqf$ & $\sqf$$\sqf$$\sqf$$\sqf$$\sqf$ & $\sqf$$\sqf$$\sqf$$\sqf$$\sqf$ & 4.88 & .33\\

Chinese Writing & $\sqf$$\sqf$$\sqf$$\sqf$$\sqf$ & $\sqf$$\sqf$$\sqf$$\sqf$$\sqf$ & $\sqf$$\sqf$$\sqf$$\sqf$$\sqe$ & $\sqf$$\sqf$$\sqf$$\sqf$$\sqf$ & $\sqf$$\sqf$$\sqf$$\sqf$$\sqf$ & $\sqf$$\sqf$$\sqf$$\sqf$$\sqf$ & $\sqf$$\sqf$$\sqf$$\sqf$$\sqf$ & $\sqf$$\sqf$$\sqf$$\sqf$$\sqf$ & 4.88 & .33\\

\bottomrule
\noalign{\vskip 3pt}
\multicolumn{11}{>{\raggedright\arraybackslash}p{\dimexpr\linewidth-2\tabcolsep\relax}}{
\footnotesize \textit{Note.} 
Self-assessment is reported on 5-point rating scales, e.g. 4/5 $\sqf\sqf\sqf\sqf\sqe$.
Expertise rating labels: strongly disagree | disagree | neutral | agree | strongly agree.
Language rating labels: Basic (A1-A2) | Intermediate (B1) | Upper Intermediate (B2) | Advanced (C1) | Mastery (C2).
Average and SD values represent numerical rating values (1-5).
} \\
\end{tabularx}
\end{table*}

\paragraph{Participants}
Four external bilingual experts with backgrounds in HCI and psychology took part in the first round, held remotely via Zoom and lasting approximately $90$ minutes. Three of them held a Master's degree and one a Doctoral degree.
The average age was $29$ years ($SD=2.55$), with one female and three male participants.
A second consolidation round was conducted by four additional native Chinese-speaking HCI researchers, three of them holding a Master's degree, and one a Bachelor's (average age $24.5$, $SD=1.12$).
Expertise levels and language skills were self-assessed using 5-point rating scales, as visualized in \autoref{tab:experts_rating_inverted}. 
All participants rated their English and Chinese reading and writing skills as advanced (C1) or mastery (C2).
Expertise in psychology, HCI, affective systems, emotion, social and behavioral theories, as well as the development and application of scientific scales varied between researchers, ensuring diverse perspectives.

\paragraph{Procedure}
The session was led by two of the authors and conducted mainly in Chinese, with occasional use of English.
After a brief introduction, each participant produced an individual translation of the original scale in order to gain new perspectives and familiarize themselves with the context.
This step took about $15$ minutes.
Together with the translations from the initial phase (\autoref{ss:translation}), these individual versions served as the base for an item-by-item discussion, which lasted around $75$ minutes.
The aim was to agree on a consolidated translation for each item that accurately reflects the contextual meaning of the original scale while taking into account the cultural and grammatical features of Chinese.
Most items were settled without major disagreement.
For item E2 the panel could not converge on a wording that captured the intended meaning of the source item, yet consolidated on one variant after extended discussion.
A back-translation of the consolidated version by two new independent translators and the AI models confirmed that uncertainty.
We therefore convened a second consolidation round with four additional native Chinese-speaking HCI researchers (see P5--P8 in \autoref{tab:experts_rating_inverted}), who settled on \cn{该系统表现出情感智能}.
Back-translation confirmed that the revised item aligned with the source scale.
The final Chinese PETS is shown in \autoref{tab:pets}.

\paragraph{Challenges}
Because Chinese does not inflect for tense, the central difficulty was keeping each item anchored to the interaction the respondent had just completed.
Chinese marks this through aspectual particles such as \cn{了、过、着} and temporal adverbials such as \cn{过去、现在、以后}, of which only the perfective \cn{了} fits an item reporting a bounded past event.
The experts therefore introduced a consistent determiner (\cn{该系统}) together with, for items reporting an event, an aspectual marker (e.g., \cn{考虑到了}, \cn{表达出了}), which anchors the episode without lengthening the items.
Three items required deliberation beyond this systematic adjustment.
For E2 (``The system seemed emotionally intelligent''), literal compounding of \emph{emotionally} and \emph{intelligent} produced unfamiliar collocations in Chinese.
The first consolidated version resolved this adverbially, but back-translation revealed that it expressed intelligence conveyed through emotion rather than emotional intelligence as such, which the reworked wording corrects by treating \cn{情感智能} as a nominal compound.
As the item attributes a property rather than reporting an event, it carries no aspectual marker.
For E4 (``The system sympathized with me''), the panel weighed \cn{同情} against \cn{共情}.
The former often implies pity in Chinese usage, whereas the latter aligns with \emph{empathy} in the sense of shared understanding, and back-translations consistently returned \emph{empathy} for \cn{共情}, so it was adopted for conceptual clarity.
For U3 (``I trusted the system''), a literal past-tense assertion sounded abrupt in Chinese, since \cn{信任} denotes a state and resists a completion reading.
The group therefore aligned on \cn{我对该系统产生了信任}, which frames trust as something that developed and thereby lets the aspectual marker apply to the onset of the state.
These examples illustrate why human judgment remains crucial for conveying appropriate tone in user-facing instruments, despite advances in machine translation.


\section{Validation}
\label{s:validation}
We designed the validation study to match the methodology of previous PETS translation and validation studies~\cite{schmidmaier2025systematic,schmidmaier2024perceived}, using identical procedures for participant recruitment, data collection, and evaluation.

\begin{figure*}[t]
    \centering
    \begin{subfigure}[b]{0.24\linewidth}
        \includegraphics[width=\linewidth]{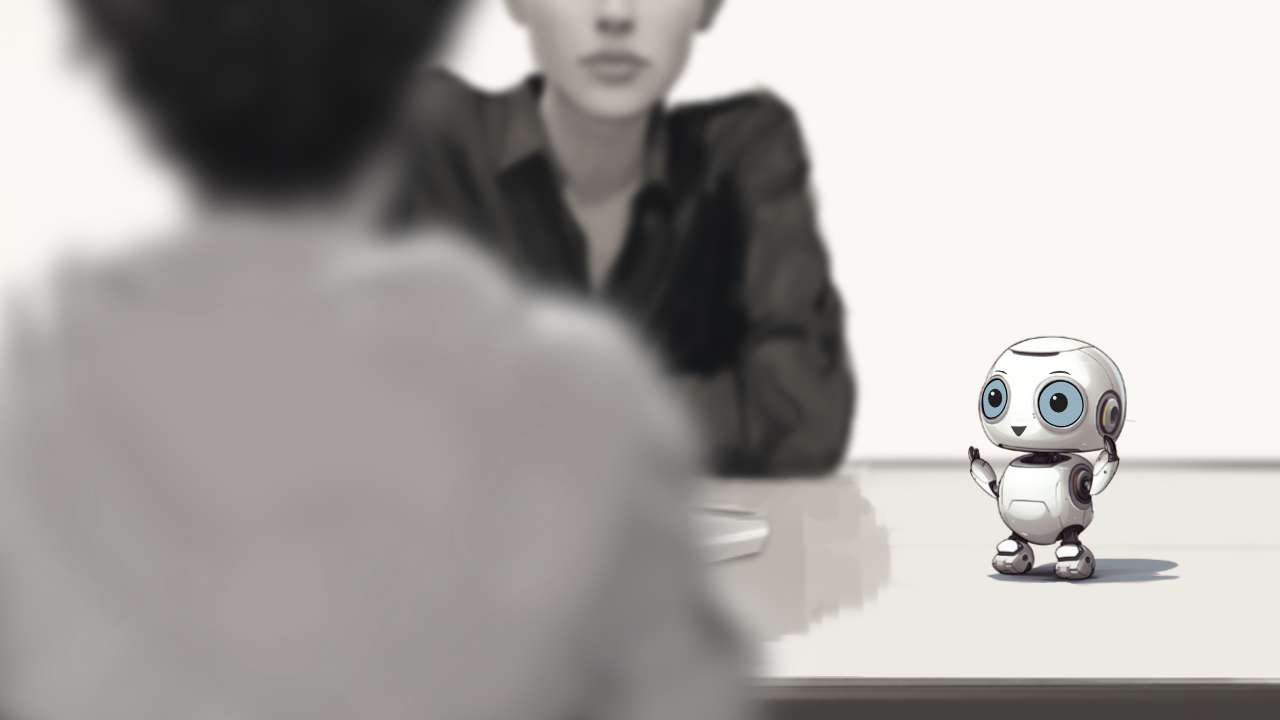}
        \caption{game companion (2:05~min)}
    \end{subfigure}
    \hfill
    \begin{subfigure}[b]{0.24\linewidth}
        \includegraphics[width=\linewidth]{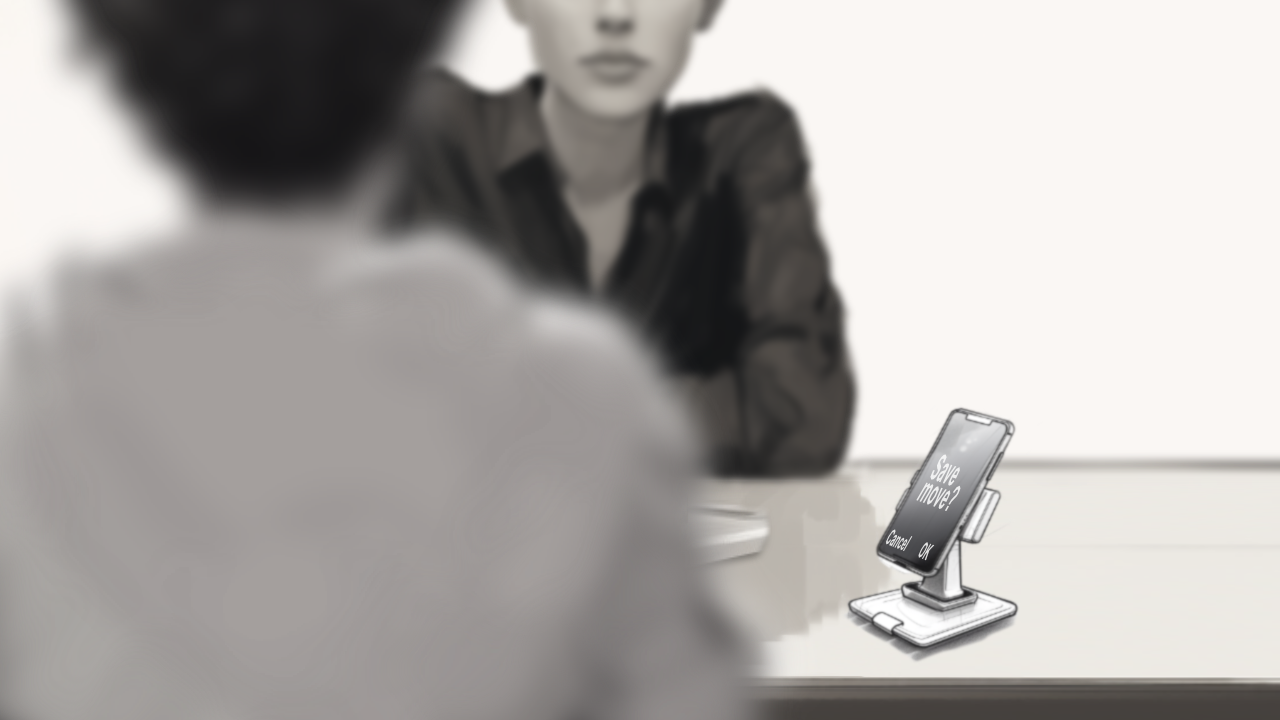}
        \caption{game app (1:34~min)}
    \end{subfigure}
    \hfill
    \begin{subfigure}[b]{0.24\linewidth}
        \includegraphics[width=\linewidth]{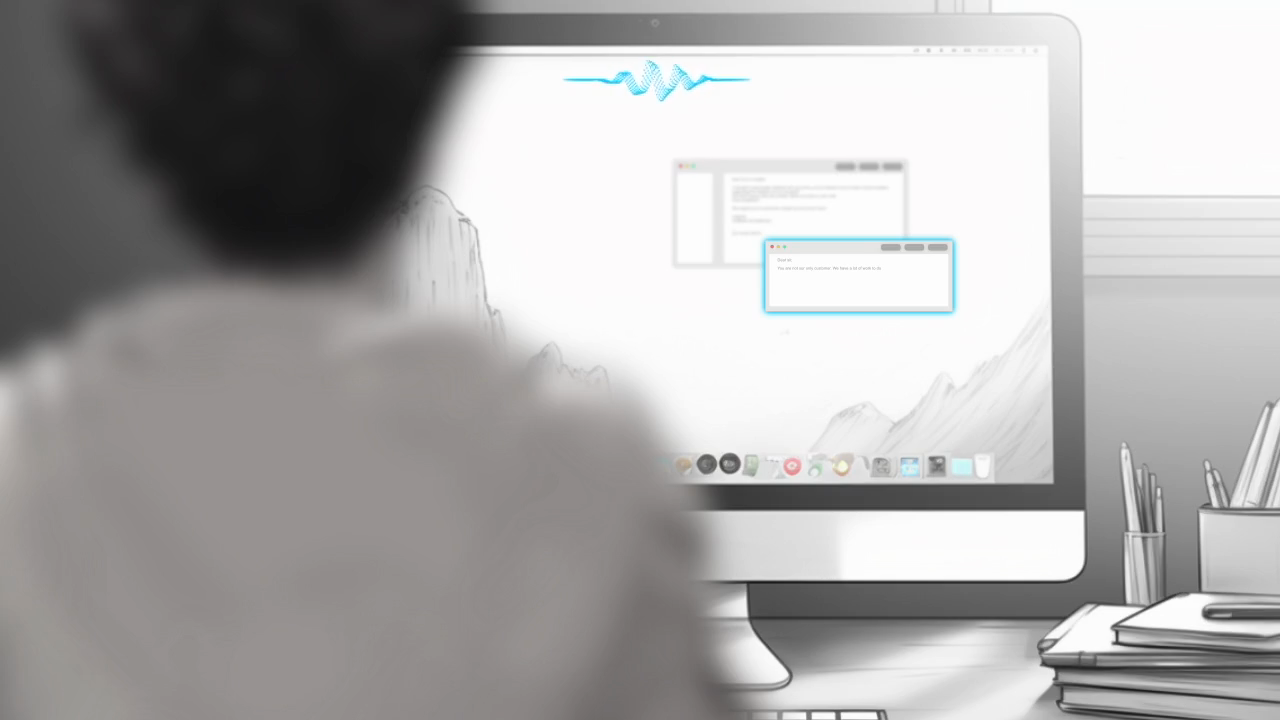}
        \caption{work companion (2:11~min)}
    \end{subfigure}
    \hfill
    \begin{subfigure}[b]{0.24\linewidth}
        \includegraphics[width=\linewidth]{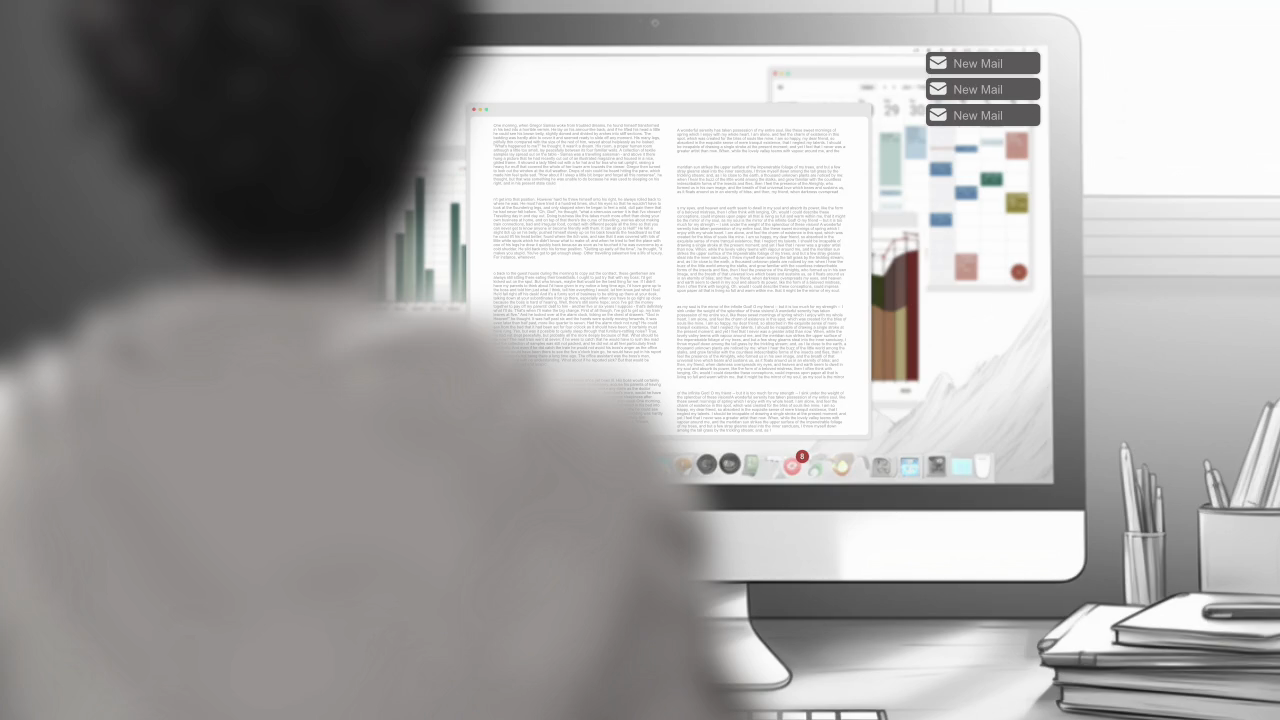}
        \caption{work app (1:33~min)}
    \end{subfigure}
    
    \caption{Screenshots of the four animated test scenarios: (a) an empathic robot companion, (b) a smartphone game application, (c) an empathic voice assistant, and (d) a notification-based office application. All scenarios included Mandarin Chinese audio tracks. }
    \label{fig:videoscenarios}
    \Description{Four images showing different human-system interaction scenarios in a study. From left to right: (a) a person facing a small robot companion across a table (2:05 min duration), (b) the same setting with a smartphone instead of robot (1:34 min), (c) a person at a computer showing work assistant interface (2:11 min), and (d) the same computer displaying a different office application interface (1:33 min). The images compare empathic vs. non-empathic interactions in both game and work contexts, all with accompanying audio narratives.}
\end{figure*}

\subsection{Participants}
\label{ss:study_participants}
For the Chinese sample, we recruited $N=200$ new participants, ensuring sufficient power according to confirmatory factor analysis guidelines~\cite{Chen2007-cv,Wolf2013-ad}.
Together with the preexisting English validation sample, this yielded a total of $N=400$ participants used for validation.

\paragraph{Chinese Sample}
The 200 native Chinese-speaking participants all resided in mainland China, and were recruited through the commercial panel provider \textit{dataSpring}.
The mean age of the participants was $35.3$ years ($SD=11.0$) with $105$ participants identifying as female and $95$ as male.

\paragraph{English Sample}
The English-speaking sample ($N=200$) was the same as used by \citet{schmidmaier2025systematic}, and is publicly available on \url{https://perceived-empathy-of-technology-scale.com}.
Participants were recruited via \textit{Prolific}, and resided in $20$ different countries, with the majority of $167$ in the European Economic Area, $29$ in North America, two in the Asia-Pacific region, and one each from the Middle East and South America.
Average age was $35.2$ years ($SD=11.8$), with $95$ participants identifying as female, $104$ as male, and one as non-binary.

\subsection{Material}
\label{ss:study_material}
For the validation study, we re-used the four original task videos from the PETS publication~\cite{schmidmaier2024perceived}.
These video scenarios showed interaction with four different systems.
Two systems, a game companion robot (a) and a work companion application (c), were designed to express empathic behavior, for example, by expressing understanding and emotional expressions.
The other two systems, a game support application (b) and an office work application (d), were designed as non-empathic counterparts, displaying a more task-like interface.
\autoref{fig:videoscenarios} shows a screenshot for each video scenario.
For translation, first a Chinese-speaking member of our team translated the existing voice lines with the help of AI translation tools (DeepL), ensuring natural phrasing.
Then we generated Mandarin voice tracks with \href{https://murf.ai}{Murf AI}.
Since Mandarin narration is typically longer than English, we padded video frames to maintain synchronization without altering event timing. 
Voices were selected to match the character of the original videos and the role of each system in the scenario.
For example, we used ``Liu'' (female) for the narrator, ``Chui'' (female) for the robot in scenario (a), and ``Jiao'' (female) for the companion in scenario (c).
This approach preserved the original scenarios' empathic vs. non-empathic behavioral contrasts while ensuring linguistic naturalness and experimental control.


\subsection{Procedure}
\label{ss:study_procedure}
The study procedure mirrored previous PETS validation studies \cite{schmidmaier2024perceived}, with all instructions and tasks presented in Mandarin Chinese.
Following informed consent and study explanation, we gathered participant demographics, including age, gender, and location.
Participants were randomly allocated to one of four scenarios and required to view the full video scenario at least once.
As a comprehension check, participants had to write a scenario summary after the video task.
They then evaluated the system using the 10-item Chinese PETS scale (see \autoref{tab:pets}) with randomized item presentation.
Response sliders had an internal range from $0$ to $100$, and were labeled \cn{非常不同意} (``strongly disagree'') and \cn{非常同意} (``strongly agree'') on their ends.
Mean completion time for the \textit{Chinese} sample was $8.0$ minutes ($SD=6.8$),  similar to the mean completion time of $7.7$ minutes ($SD=4.1$) reported for the original English sample~\cite{schmidmaier2025systematic}.
Participants were recruited and compensated by \textit{dataSpring} according to their internal panel reward scheme, at a cost of $1.60$~\euro~per completed response.
The study was approved by the ethics committee of our institution. 
Participation was voluntary and could be canceled at any time without consequences. 
All participants gave informed consent after being informed about the study details and the processing of their data. Person-related data were processed only by our research team.

\begin{figure*}[t!]
    \centering
    \includegraphics[width=\linewidth]{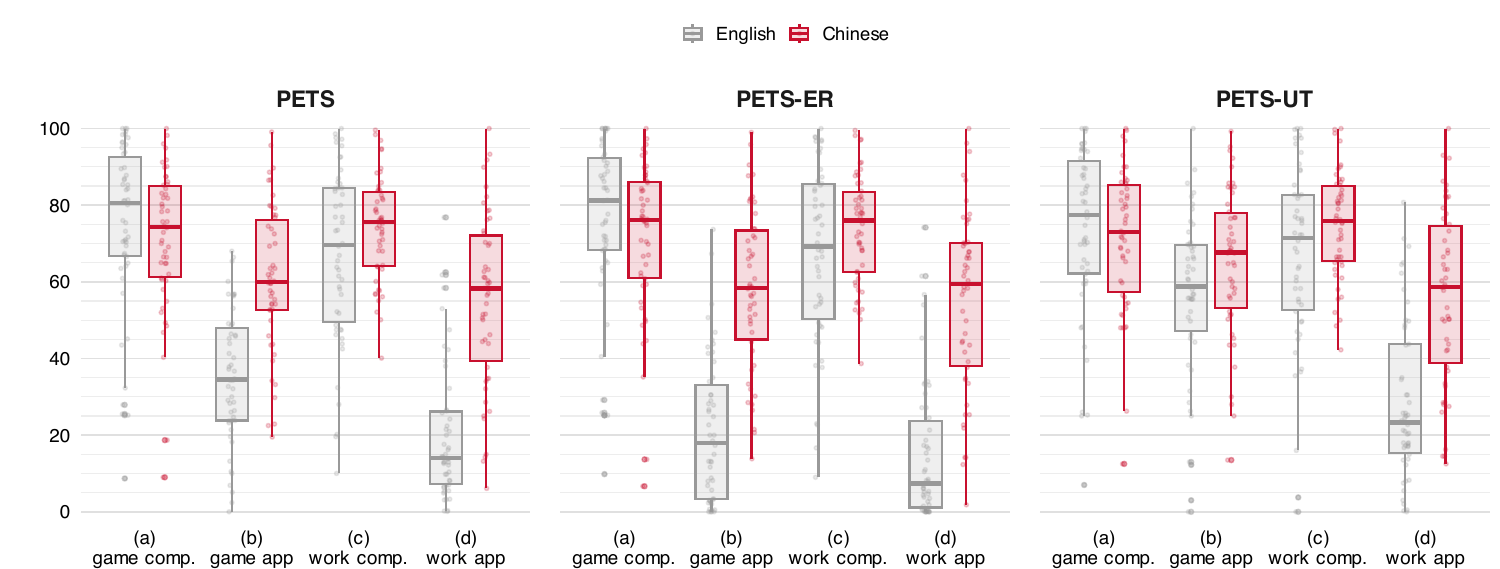}
    \caption{
    PETS, PETS-ER, and PETS-UT scores across the four scenarios for the English and the Chinese scales.
    Scenarios depicted two empathic companions (a, c) and two non-empathic applications (b, d). N=50 per scenario and language.}
    \label{fig:boxplots}
    \Description{} 
\end{figure*}

\subsection{Results}
\label{ss:results}
For analysis, we closely followed the validation process as described by \citet{Klotz2023-ww} and applied by \citet{schmidmaier2025systematic}, assessing the equivalence between the original English and the translated Chinese PETS.
Median PETS ratings are depicted in \autoref{fig:boxplots}.

\paragraph{PETS Scenario Effects within Samples}
Shapiro-Wilk tests on the ANOVA residuals indicated departures from normality for all three measures in the English sample ($p<.05$) and for PETS-ER in the Chinese sample ($p=.020$), so we used non-parametric Kruskal-Wallis tests throughout, consistent with the original English PETS validation.
Scenario differences reached significance in both samples for overall PETS scores as well as for PETS-ER and PETS-UT (all $p<.001$).
Dunn's post hoc tests with Bonferroni correction set the empathic scenarios (a, c) apart from the non-empathic ones (b, d) on overall PETS and PETS-ER across all four cross-condition comparisons ($p<.05$), which confirms that the intended empathy manipulation transferred to the Chinese version (\autoref{tab:withinsamples_posthoc}).
PETS-UT differentiated less clearly, not separating the game app (b) from the game companion (a) ($p=.306$), and only marginally from the work companion (c) ($p=.050$), where the latter pairing had also been non-significant in the English data.
As expected, the two empathic scenarios (a vs. c) did not differ from each other, and neither did the two non-empathic ones (b vs. d).

\begin{table}[b]
\centering
\small
\caption{Bayesian Mann-Whitney U tests (data augmentation, 5 chains of 10000 iterations) over combined data from both samples, examining language group effects for each scenario.}
\label{tab:betweensamples_bayesian}
\begin{tabularx}{\linewidth}{Xld{6.3}d{4.3}d{1.3}}
\toprule
\textbf{Measure}
& \textbf{Scenario}
& \multicolumn{1}{c}{\textbf{$BF_{10}$}}
& \multicolumn{1}{c}{\textbf{W}}
& \multicolumn{1}{c}{\textbf{Rhat}}
\\
\midrule
PETS & (a) game companion & .480 & 1454.500 & 1.000 \\
PETS-ER & (a) game companion & .635 & 1472.000 & 1.000 \\
PETS-UT & (a) game companion & .278 & 1386.000 & 1.000 \\
\midrule
PETS & (b) game app & \multicolumn{1}{r}{$>$1000} & 422.000 & 1.009 \\
PETS-ER & (b) game app & \multicolumn{1}{r}{$>$1000} & 254.500 & 1.002 \\
PETS-UT & (b) game app & 1.462 & 957.000 & 1.000 \\
\midrule
PETS & (c) work companion & .436 & 1073.500 & 1.000 \\
PETS-ER & (c) work companion & .459 & 1078.000 & 1.000 \\
PETS-UT & (c) work companion & .572 & 1030.500 & 1.000 \\
\midrule
PETS & (d) work app & \multicolumn{1}{r}{$>$1000} & 307.000 & 1.003 \\
PETS-ER & (d) work app & \multicolumn{1}{r}{$>$1000} & 236.500 & 1.007 \\
PETS-UT & (d) work app & \multicolumn{1}{r}{$>$1000} & 476.500 & 1.005 \\
\bottomrule
\noalign{\vskip 3pt}
\multicolumn{5}{>{\raggedright\arraybackslash}p{\dimexpr\linewidth-2\tabcolsep\relax}}{\footnotesize \textit{Note.} Bayes factors above $1000$ are reported as $>$1000, since values in this range are estimated imprecisely and all indicate extreme evidence.} \\
\end{tabularx}
\end{table}


\paragraph{PETS Language Effects between Samples}
Kruskal-Wallis tests on the combined data indicated significant language-group differences for PETS, PETS-ER, and PETS-UT (all $p<.001$), alongside the expected scenario effects.
Mann-Whitney U tests (\autoref{tab:betweensamples_posthoc}) located the differences exclusively in the non-empathic conditions.
As also visible in \autoref{fig:boxplots}, Chinese participants rated the game app and the work app significantly higher on all three measures ($p<.05$), with median differences reaching $52$ points on PETS-ER.
The companion scenarios (a, c) did not differ on any measure (all $p>.12$).
We additionally ran Bayesian Mann-Whitney U tests (data augmentation, 5 chains of 10000 iterations) in JASP~\cite{JASP2025} to quantify the evidence for and against a language effect in each scenario, with all chains converging ($\widehat{R} \leq 1.009$).
For the empathic scenarios, the Bayes factors favored the null throughout ($BF_{10} = .278$--$.635$, \autoref{tab:betweensamples_bayesian}), giving anecdotal evidence against a language effect in five of six cases and moderate evidence for PETS-UT in the game companion scenario.
The absence of significant differences there can thus be read as tentative support for equivalence.
For the non-empathic scenarios, the evidence for a language effect was extreme on overall PETS and PETS-ER in both the game app and the work app, and likewise on PETS-UT in the work app.
Only PETS-UT in the game app remained anecdotal ($BF_{10} = 1.462$), which qualifies the significant frequentist result obtained there ($p=.044$).
Language effects were therefore confined to the non-empathic scenarios, where the Chinese sample attributed considerably more perceived empathy to the agents than the English sample did. 

\begin{figure}[t]
    \centering
    \includegraphics[width=.48\linewidth]{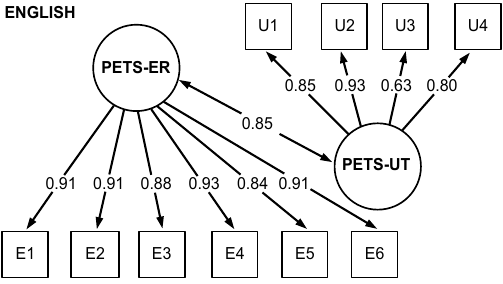}
    \hfill
    \includegraphics[width=.48\linewidth]{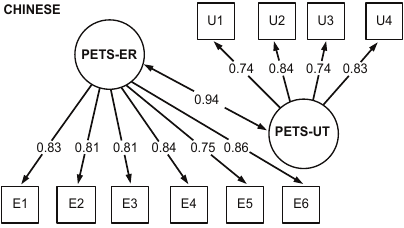}
    \caption{Factor structure and standardized loadings of the English PETS and the translated Chinese version.}
    \Description{}
    \label{fig:pets_factors}
\end{figure}

\paragraph{Invariance}
Following established guidelines~\cite{Hu1999-uk,Chen2007-cv}, we ran a CFA in the \textit{lavaan} package in R~\cite{Rosseel-lavaan} and examined configural, metric, and scalar invariance between the English and Chinese PETS versions.
Fit for the \textit{configural invariance} model was acceptable ($CFI=.974$, $TLI=.966$, $SRMR=.031$), with only $RMSEA=.082$ falling marginally outside the recommended cutoff (\autoref{tab:cfa_results}).
The two-factor structure therefore appears to hold in both the English and Chinese scales.
Adding equality constraints on the factor loadings for \textit{metric invariance} left fit essentially unchanged ($CFI=.971$, $TLI=.966$, $SRMR=.055$, $RMSEA=.081$, $\Delta CFI=-.003$, $\Delta RMSEA=-.001$).
Item loadings can thus be treated as comparable across groups, pointing to functional equivalence of the two constructs.
Standardized loadings were strong throughout (\autoref{fig:pets_factors}), but the ER items were uniformly attenuated in the Chinese version ($.75$--$.86$ against $.84$--$.93$ in English).
Within UT, item U3 loaded considerably higher and item U1 somewhat lower than in the English baseline, so the factor rests more evenly on its four items, a pattern we return to in \autoref{ss:discussion_pets_ut}.
Adding intercept constraints for \textit{scalar invariance} again yielded acceptable fit ($CFI=.964$, $TLI=.962$, $SRMR=.061$, $RMSEA=.086$) with only modest deterioration ($\Delta CFI=-.007$, $\Delta RMSEA=.005$).
Loadings and intercepts can therefore be regarded as equivalent, giving reasonable grounds for latent mean comparisons across the two versions.

\begin{table*}[t]
\centering
\caption{
Invariance testing results comparing the English PETS to the Chinese version. 
Model comparisons show the difference between metric and configural as well as scalar and metric model results.
}
\label{tab:cfa_results}
\begin{tabularx}{\linewidth}{Xcccccccccc}
\toprule
& \multicolumn{6}{c}{\textbf{Model Fit Indices}} & \multicolumn{4}{c}{\textbf{Model Comparisons}} \\
\cmidrule(lr){2-7} \cmidrule(lr){8-11}
Model & $\chi^2$ & df & CFI & TLI & RMSEA & SRMR & $\Delta\chi^2$ & $\Delta$df & $\Delta$CFI & $\Delta$RMSEA \\
\midrule
1. Configural Invariance 
& 158.84 & 68 & .974 & .966 & .082 & .031 
& --- & --- & --- & --- \\
2. Metric Invariance 
& 175.82 & 76 & .971 & .966 & .081 & .055 
& 16.98 & 8 & -.003 & -.001 \\
3. Scalar Invariance 
& 208.64 & 84 & .964 & .962 & .086 & .061 
& 32.82 & 8 & -.007 & .005 \\
\bottomrule
\end{tabularx}
\end{table*}


\paragraph{Internal Consistency}
Internal consistency was assessed via Cronbach's alpha for the full scale and for each subscale in both languages~\cite{George2024-ag}.
Because alpha rests on the assumption of equal item loadings, we report McDonald's omega alongside it, as omega is derived from the loadings themselves and thus fits the two-factor structure of the PETS more closely.
Reliability of the Chinese PETS was excellent overall ($\alpha = .944$, $\omega = .949$), falling marginally short of the original English version ($\alpha = .956$, $\omega = .965$).
On the PETS-ER subscale both versions again reached excellent values ($\alpha = .960$, $\omega = .960$ in English against $\alpha = .922$, $\omega = .922$ in Chinese), the largest reduction observed across the three scales.
The PETS-UT subscale was good in either language ($\alpha = .877$, $\omega = .882$ in English, $\alpha = .863$, $\omega = .866$ in Chinese).
Throughout, omega deviated from the corresponding alpha by no more than $.01$, so the equal-loading assumption had no meaningful bearing on the estimates.

\section{Discussion}
\label{s:discussion}
In this section we discuss the psychometric equivalence of the Chinese PETS, the divergence between the samples in the non-empathic scenarios, the item-level differences within the PETS-UT subscale, and the limitations of our design.

\subsection{Translation Equivalence and Validity}
\label{ss:discussion_validity}
As reported in \autoref{ss:results}, the invariance tests (\autoref{tab:cfa_results}) support the Chinese PETS as measuring the same constructs as the English original, since constraining loadings and intercepts to equality did not meaningfully worsen model fit, so scores from the two versions can be meaningfully compared.
Similar to the existing German PETS translation~\cite{schmidmaier2025systematic}, our RMSEA values slightly exceeded the ideal threshold of $.08$~\cite{Hu1999-uk}, reaching at most $.086$.
In this context, \citet{schmidmaier2025systematic} refer to \citet{Kenny2003-hn}, who note that RMSEA tends to penalize simpler models with fewer degrees of freedom, artificially inflating values and potentially leading to rejection of properly specified models.
Given that TLI and CFI stayed within their recommended ranges and that internal consistency was strong across the full scale and both subscales, we regard the elevated RMSEA as a property of the model specification rather than evidence against the translation.

\paragraph{Subscale Separability}
The two factors correlated more strongly in the Chinese sample ($r = .94$) than in the English one ($r = .85$), which suggests that the two dimensions were perceived as more closely related.
Since correlations of this size can make subscales difficult to tell apart empirically, we compared the two-factor model against a single-factor alternative.
The two-factor solution fitted significantly better in both samples, for Chinese ($\Delta\chi^2(1) = 14.93$, $p<.001$) and for English participants ($\Delta\chi^2(1) = 117.91$, $p<.001$), so the distinction is supported in each case.
The margin, however, differed considerably, as moving to a single factor cost the English model $.058$ in CFI and $.069$ in RMSEA, against only $.009$ and $.011$ in the Chinese model.
PETS-ER and PETS-UT thus remain distinguishable in Chinese, though subscale scores may be read as closely related aspects of a common impression rather than as independent dimensions.

\medskip
\noindent
Based on these findings, we consider the Chinese PETS a psychometrically sound instrument for evaluating perceived empathy in interactive systems among Chinese-speaking populations, facilitating cross-cultural comparative studies.

\subsection{Cross-Cultural Scenario Effects}
\label{ss:discussion_language_effects}
Beyond structural equivalence, the behavioral data revealed a systematic difference between the samples that was confined to one type of scenario.
Ratings of the empathic scenarios (a, c) were statistically indistinguishable across language groups, with Bayes factors favoring the absence of a language effect throughout ($BF_{10} \leq .635$).
For the non-empathic scenarios (b, d), by contrast, the evidence for a language effect was extreme on overall PETS and PETS-ER ($BF_{10} > 1000$ in both), as Chinese participants assigned markedly higher scores than the English sample (\autoref{fig:boxplots}).
A general translation artifact would shift ratings uniformly across all four scenarios, so this asymmetry points instead to differences in how the two groups evaluated the absence of empathic behavior.
The same pattern appeared within the Chinese sample, where PETS-UT no longer separated the game app (b) from either companion scenario, indicating that the empathy manipulation registered less sharply on the understanding and trust items than in the English data.
In the following, we outline three potential explanations consistent with this pattern.

\paragraph{Response Style}
The asymmetry matches a documented cross-cultural response style.
East Asian respondents use scale midpoints more and endpoints less than North American respondents~\cite{chen1995response}, and the difference is not symmetric, as Chinese respondents show weaker extreme \textit{rejecting} responses than British respondents while showing no difference at the affirming end~\cite{guo2019cross}.
Our non-empathic scenarios draw low ratings and our empathic ones high ratings, so a weaker tendency towards extreme low ratings predicts precisely what we observe.

\paragraph{Recruitment Confound}
The samples were recruited through different channels.
The English baseline came from Prolific, whereas the Chinese sample was recruited via \textit{dataSpring}, due to a lack of mainland Chinese participants on Prolific.
Since data quality has been reported to vary across recruitment sources~\cite{peer2022data,douglas2023data}, recruitment channel and language group remain confounded in our comparison.

\paragraph{Cultural Interpretation}
The divergence may be cultural in origin.
If the absence of empathic behavior constitutes a weaker and more interpretively open signal than its presence, judgments of non-empathic systems would draw more heavily on culturally shaped expectations, producing an asymmetry of the observed form.
The greater sensitivity of PETS-ER compared to PETS-UT fits this reading, as emotional responses are plausibly more tied to culturally shaped empathy conceptions than functional judgments of understanding and trust.
This account also yields a testable prediction, namely that with response style and recruitment controlled, the divergence should persist for the non-empathic scenarios and remain absent for the empathic ones.

\medskip
\noindent
As these accounts are not mutually exclusive, we present them as candidates that future work with matched recruitment and response style corrections could separate.

\subsection{Cross-Cultural Item Effects}
\label{ss:discussion_pets_ut}
While the internal consistency of the Chinese PETS-UT subscale was comparable to the English version ($\alpha = .863$ vs. $\alpha = .877$), the loadings within the subscale were distributed differently.
Item U3 (\cn{我对该系统产生了信任。}, ``I trusted the system'') loaded considerably higher in the Chinese version ($.74$) than in the English one ($.63$), while item U1 dropped from $.85$ to $.74$, so the factor rests more evenly across its four items rather than being carried primarily by a single indicator.
This pattern might indicate culturally specific interpretations of system understanding and trust, potentially reflecting the importance of guanxi (\cn{关系}) and relational trust in Chinese culture.
We speculate that Chinese participants might perceive system trustworthiness more holistically, possibly integrating relational expectations with functional performance, which could result in stronger cohesion of the trust item within PETS-UT.
The stronger correlation between the two factors in the Chinese sample (\autoref{ss:discussion_validity}) fits this reading, as it suggests that emotional responsiveness and understanding and trust were judged as facets of a single overall impression rather than as separable qualities.
Read alongside the cultural interpretation offered in \autoref{ss:discussion_language_effects}, these observations point in the same direction while resting on different evidence, as that account concerned differences in mean ratings whereas the present one concerns the internal structure of the scale.

\subsection{Limitations}
\label{ss:discussion_limitations}
Our findings should be read with the following constraints in mind.
We established factorial validity but not convergent validity, as PETS scores were never compared against dispositional empathy or other related measures.
Language group and recruitment source vary together across our samples, and we collected no measure of scale-use tendencies, so response style may contribute to the group effects described in \autoref{ss:discussion_language_effects}.
We therefore treat the reported language effects as an upper bound on cross-cultural difference.
Participants observed the systems rather than using them, as in earlier PETS work~\cite{schmidmaier2024perceived,schmidmaier2025systematic}, which potentially weakens ecological validity.
The scenarios were also carried over from the English validation rather than built around Chinese applications, and may misrepresent how empathic features appear in technologies widespread in China.
RMSEA stayed slightly above the recommended threshold in all models, and while \citet{Kenny2003-hn} attribute such inflation to models with few degrees of freedom, bifactor modeling~\cite{Reise2012-zo} or exploratory structural equation modeling~\cite{Marsh2014-kg} would put the factor structure on firmer ground.

\subsection{Future Directions}
\label{ss:future}
Several of the constraints above point directly to next steps.
Validating the Chinese PETS against established trait measures would extend the evidence from factorial to convergent validity, while interactive rather than observational designs and scenarios built around Chinese applications would test the scale where it is meant to be used.
Our results also raise a question beyond the instrument itself.
The divergence between samples was confined to the non-empathic scenarios, which suggests that cultures may differ less in how they recognize empathic behavior than in how they read its absence.
Testing that proposition requires designs that vary empathic behavior along a continuum rather than as a binary contrast, together with matched recruitment so that cultural differences can be separated from sampling ones.
Should the asymmetry hold, it would carry a practical consequence, namely that the risk of an empathic system failing to appear empathic is unevenly distributed across cultural contexts.
More broadly, the validated scale opens up empirical work in a context where empathic systems are deployed at scale, from social robots and voice assistants to customer service and public administration, including questions about their consequences, such as whether empathic agents alleviate loneliness or deepen engagement, and at what cost.

\section{Conclusion}
We present a validated Chinese translation of the Perceived Empathy of Technology Scale (PETS), targeting Putonghua (\cn{普通话}) and Simplified Chinese (\cn{简体中文}), following the back-translation and validation procedures of previous PETS translations.
Comparing $N=200$ Chinese-speaking participants against an English sample of equal size, CFA and invariance testing confirmed that the two-factor structure, item loadings, and intercepts transfer to the translated version, with high reliability throughout.
Ratings of empathic system behavior were equivalent across languages, while non-empathic behavior was rated substantially higher by Chinese participants, suggesting that the absence of empathy is read less consistently across cultures than its presence.
The Chinese PETS thus provides a sound basis for assessing perceived empathy in artificial systems and for the cross-cultural comparisons such assessment invites.

\section*{Author Contributions}
\textbf{Matthias Schmidmaier}: Conceptualization, Data curation, Formal Analysis, Investigation, Methodology, Project administration, Resources, Software, Validation, Visualization, Writing – original draft, Writing – review \& editing; \textbf{Jingyao Zheng}: Investigation, Resources, Writing – original draft, Writing – review \& editing; \textbf{Jie Song}: Data curation, Investigation, Methodology, Writing – original draft; \textbf{Lik-Hang Lee}: Resources, Supervision; \textbf{Sven Mayer}: Conceptualization, Formal Analysis, Funding acquisition, Investigation, Resources, Supervision, Writing – original draft, Writing – review \& editing

\begin{acks}
This work has been partly supported by the Research Center Trustworthy Data Science and Security (\href{https://rc-trust.ai}{https://rc-trust.ai}), one of the Research Alliance centers within the UA Ruhr (\href{https://uaruhr.de}{https://uaruhr.de}).
Portions of the text in this paper were refined with the assistance of generative AI tools, which were used solely for phrasing and language optimization. All ideas, analyses, and conclusions are the authors’ own, and the authors take full responsibility for the content.
\end{acks}

\section*{Use of AI}
Portions of the text in this paper were refined with the assistance of generative AI tools, which were used for phrasing and language optimization. 
All ideas, analyses, and conclusions are the authors' own.
The use of AI for creation of study materials or translations is mentioned in the respective sections.

\bibliographystyle{ACM-Reference-Format}
\bibliography{bibliography}

\clearpage

\appendix

\section{Appendix}

\begin{table}[H]
\centering
\caption{Post-hoc Dunn's test with Bonferroni correction between scenarios within each independent language sample.}
\label{tab:withinsamples_posthoc}
\small
\begin{tabularx}{\linewidth}{Xlcd{1.3}d{1.3}c}
\toprule
\textbf{Comparison} & \textbf{Measure} & \textbf{Language} & \multicolumn{1}{c}{\textbf{Z}} & \multicolumn{1}{c}{\textbf{p-value}} & \\
\midrule
(b) game app vs. (a) game companion & PETS & English & -6.46 & <.001 & *** \\
(b) game app vs. (a) game companion & PETS-ER & English & -7.75 & <.001 & *** \\
(b) game app vs. (a) game companion & PETS-UT & English & -3.37 & .002 & ** \\
(b) game app vs. (a) game companion & PETS & Chinese & -2.95 & .010 & ** \\
(b) game app vs. (a) game companion & PETS-ER & Chinese & -3.38 & .002 & ** \\
(b) game app vs. (a) game companion & PETS-UT & Chinese & -1.64 & .306 &  \\
\midrule
(b) game app vs. (d) work app & PETS & English & 2.59 & .029 & * \\
(b) game app vs. (d) work app & PETS-ER & English & 1.12 & .787 &  \\
(b) game app vs. (d) work app & PETS-UT & English & 4.51 & <.001 & *** \\
(b) game app vs. (d) work app & PETS & Chinese & .91 & 1.000 &  \\
(b) game app vs. (d) work app & PETS-ER & Chinese & .47 & 1.000 &  \\
(b) game app vs. (d) work app & PETS-UT & Chinese & 1.86 & .191 &  \\
\midrule
(b) game app vs. (c) work companion & PETS & English & -5.19 & <.001 & *** \\
(b) game app vs. (c) work companion & PETS-ER & English & -6.52 & <.001 & *** \\
(b) game app vs. (c) work companion & PETS-UT & English & -2.05 & .122 &  \\
(b) game app vs. (c) work companion & PETS & Chinese & -3.38 & .002 & ** \\
(b) game app vs. (c) work companion & PETS-ER & Chinese & -3.66 & <.001 & *** \\
(b) game app vs. (c) work companion & PETS-UT & Chinese & -2.39 & .050 &  \\
\midrule
(a) game companion vs. (d) work app & PETS & English & 9.05 & <.001 & *** \\
(a) game companion vs. (d) work app & PETS-ER & English & 8.87 & <.001 & *** \\
(a) game companion vs. (d) work app & PETS-UT & English & 7.88 & <.001 & *** \\
(a) game companion vs. (d) work app & PETS & Chinese & 3.86 & <.001 & *** \\
(a) game companion vs. (d) work app & PETS-ER & Chinese & 3.84 & <.001 & *** \\
(a) game companion vs. (d) work app & PETS-UT & Chinese & 3.49 & .001 & ** \\
\midrule
(a) game companion vs. (c) work companion & PETS & English & 1.27 & .612 &  \\
(a) game companion vs. (c) work companion & PETS-ER & English & 1.23 & .659 &  \\
(a) game companion vs. (c) work companion & PETS-UT & English & 1.33 & .554 &  \\
(a) game companion vs. (c) work companion & PETS & Chinese & -.44 & 1.000 &  \\
(a) game companion vs. (c) work companion & PETS-ER & Chinese & -.29 & 1.000 &  \\
(a) game companion vs. (c) work companion & PETS-UT & Chinese & -.76 & 1.000 &  \\
\midrule
(d) work app vs. (c) work companion & PETS & English & -7.78 & <.001 & *** \\
(d) work app vs. (c) work companion & PETS-ER & English & -7.64 & <.001 & *** \\
(d) work app vs. (c) work companion & PETS-UT & English & -6.56 & <.001 & *** \\
(d) work app vs. (c) work companion & PETS & Chinese & -4.29 & <.001 & *** \\
(d) work app vs. (c) work companion & PETS-ER & Chinese & -4.13 & <.001 & *** \\
(d) work app vs. (c) work companion & PETS-UT & Chinese & -4.25 & <.001 & *** \\
\bottomrule
\multicolumn{6}{l}{\footnotesize \textit{Note.} (a) game companion, (b) game app, (c) work companion, (d) work app} \\
\end{tabularx}
\end{table}


\begin{table}[t]
\centering
\caption{Post-hoc Mann-Whitney U tests comparing the English and Chinese samples within each scenario. Negative differences indicate higher ratings in the Chinese sample.}
\label{tab:betweensamples_posthoc}
\small
\begin{tabularx}{\linewidth}{Xld{2.2}d{2.2}d{3.2}d{4.2}d{1.3}c}
\toprule
\textbf{Measure} & \textbf{Scenario} & \multicolumn{1}{c}{\textbf{Mdn\textsubscript{EN}}} & \multicolumn{1}{c}{\textbf{Mdn\textsubscript{Ch}}} & \multicolumn{1}{c}{\textbf{$\Delta$Mdn}} & \multicolumn{1}{c}{\textbf{U}} & \multicolumn{1}{c}{\textbf{p-value}} & \\
\midrule
PETS & (a) game companion & 80.60 & 74.35 & 6.25 & 1454.50 & .160 &  \\
PETS-ER & (a) game companion & 81.25 & 76.08 & 5.17 & 1472.00 & .127 &  \\
PETS-UT & (a) game companion & 77.50 & 73.00 & 4.50 & 1386.00 & .350 &  \\
\midrule
PETS & (b) game app & 34.50 & 59.95 & -25.45 & 422.00 & <.001 & *** \\
PETS-ER & (b) game app & 17.92 & 58.42 & -40.50 & 254.50 & <.001 & *** \\
PETS-UT & (b) game app & 58.75 & 67.62 & -8.88 & 957.00 & .044 & * \\
\midrule
PETS & (c) work companion & 69.65 & 75.65 & -6.00 & 1073.50 & .225 &  \\
PETS-ER & (c) work companion & 69.25 & 76.00 & -6.75 & 1078.00 & .237 &  \\
PETS-UT & (c) work companion & 71.50 & 75.88 & -4.38 & 1030.50 & .131 &  \\
\midrule
PETS & (d) work app & 14.10 & 58.25 & -44.15 & 307.00 & <.001 & *** \\
PETS-ER & (d) work app & 7.42 & 59.50 & -52.08 & 236.50 & <.001 & *** \\
PETS-UT & (d) work app & 23.25 & 58.62 & -35.38 & 476.50 & <.001 & *** \\
\bottomrule
\end{tabularx}
\end{table}


\end{document}